\documentclass[twocolumn,showpacs,preprintnumbers,amsmath,amssymb]{revtex4}

\usepackage{amssymb}
\usepackage{amsmath}
\usepackage{graphicx}
\usepackage{dcolumn}
\usepackage{bm}

\begin{document}

\preprint{APS/123-QED}
\title{Long range crossed Andreev reflections in high Tc superconductors}
\author{William J. Herrera}
\affiliation{Departamento de F\'{\i}sica, Universidad Nacional de Colombia, Bogot\'{a}%
, Colombia.}
\author{A. Levy Yeyati}
\author{A. Martin-Rodero}
\affiliation{Departamento de F\'{\i}sica Te\'orica de la Materia Condensada, Universidad Aut\'onoma
de Madrid, E-28049 Madrid, Spain.\\
}
\date{\today }

\begin{abstract}
We analyze the non-local transport properties of a $d$-wave superconductor
coupled to metallic electrodes at nanoscale distances. We show that the non-local conductance
exhibits an algebraical decay with distance rather than the exponential behavior which is found
in conventional superconductors. Crossed Andreev
processes, associated with electronic entanglement, are favored for certain 
orientations of the symmetry axes of the
superconductor with respect to the leads. These properties would allow its
experimental detection using present technologies.
\end{abstract}

\maketitle



\section{Introduction}
\begin{figure}
\begin{center}
\includegraphics[scale=0.8]{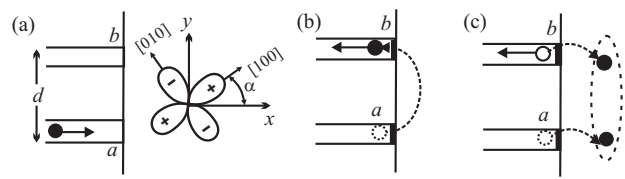}
\end{center}
\caption{Two leads on a $d$-wave superconductor. The distance between the
leads is $d$, $\protect\alpha$ is the angle between the crystalographic axes
of the superconductor and the direction normal to the surface. An incoming
electron from lead $a$ can be reflected in lead $b$ as: (b) an electron or,
(c) a hole while a Cooper pair is created in the superconductor.}
\label{Fig:1}
\end{figure}

Cooper pairs in superconducting nanostructures provide
a potential source of entangled electrons \cite{Recher1,
Chtchelkatchev2002,Bena2002,Recher2,Samuelsson2003}, a possibility that has
been recently explored in conventional superconductors both theoretically 
\cite{Byers1995,Feinberg2000,Falci2001,Feinberg2003, Melin2003a,
Maekawa2003,Melin2004,Prada2004,Brinkman2006,Alfredo2007,Kalenkov2007} and
experimentally \cite{Beckmann2004, Russo2005,
Chand2006,Beckmann2007}. In a typical experimental device, a
superconducting region is contacted by several metallic electrodes at
nanoscale distances with the aim of analyzing the non-local transport
properties at subgap voltages. In the limit of vanishing contact
transparency the non-local conductance is controlled by two type of
processes yielding opposite contributions: direct elastic tunneling of
electrons between two separate leads (elastic cotunneling, EC) and crossed
Andreev reflection (CAR) processes in which injected electrons from one lead
are reflected as holes in the other lead (see Fig. \ref{Fig:1}). The time
reverse of these last processes involve entangled electron pairs on two
separate leads \cite{comment}. In conventional superconductors the average conductance
tends to cancel due to the opposite contribution of EC and CAR processes 
\cite{Falci2001,Feinberg2003}. Several mechanisms have been proposed to
avoid a complete cancellation and have been invoked to explain the available
experimental results. Among them one can quote the use of ferromagnetic
leads \cite{Feinberg2000,Maekawa2003,Melin2003a,Melin2004} and the effect of
electron-electron interactions in certain experimental geometries \cite%
{Alfredo2007}. On the other hand, the magnitude of these non-local processes
decays exponentially with the distance between the leads on a scale fixed by
the superconducting coherence length $\xi_{0}.$ In practice this means that
non-local effects can be observed on distances of the order 10 nm to 1$\mu$m, depending on the material \cite{Beckmann2004,Russo2005}.

These effects have been much less explored in the 
case of unconventional high critical temperature  superconductors (HTcS)\cite%
{Byers1995,Melin2003b,Maekawa2006}. The characteristic small values of the
coherence length in these systems cast doubts about the observability of
non-local correlations. However, due to the anisotropy of the pair
potential, the coherence length along certain directions can be much larger
than $\xi _{0}$. This anisotropy is behind the non-local nature of the
electromagnetic response of HTcS\cite{Kosztin,Li}. In fact, some indirect
evidence of CAR\ \ processes in HTcS coupled to ferromagnetic leads has been
presented \cite{Koren2005, Koren2006}.

In the present work we analyze the non-local transport in $d$-wave
superconductors and show that in contrast to the conventional $s$-wave case,
CAR\ processes are long ranged. Moreover, we show that for certain
orientations of the axes of the superconductor with respect to the contacts,
CAR processes dominate over EC at low voltages and small contact
transparency. We believe that these findings open the possibility of using
HTcS as a source of entangled electron pairs.

\section{Crossed Andreev reflection and elastic cotunneling in high $T_C$ superconductors}
The situation to be analyzed is illustrated in Fig. \ref{Fig:1}. We consider
a semi-infinite $d$-wave superconducting region connected to two normal
leads, denoted by $a$ and $b$, and separated a distance $d$. Our aim is
finding the current induced on lead $b$, $I_b$, when a voltage $V_a$ is
applied on lead $a$. The two processes contributing to this current are
depicted on panels (b) and (c). Being the result of the diffraction of
quasiparticles by an anisotropic pair potential, the relative
weight of the two processes will be affected by the orientation of the
superconductor symmetry axis (angle $\alpha $ in Fig. \ref{Fig:1}). In fact,
EC processes are favored for electron propagation along the nodal
lines (where the order parameter vanishes), while CAR processes reach a maximum amplitude along directions 
where the modulus of the pair potential is
maximum. Although on a spatial average the contribution of the two processes should be equal as in the case of isotropic s-wave superconductors, a dominance of one of the two can be found for specific orientations of the symmetry axis with respect to the leads. These qualitative arguments allow to understand the dominance of
CAR over EC processes for the $\alpha =0$ case ($d_{x^{2}-y^{2}}$ symmetry)
and the opposite behavior in the case of $\alpha =\pi /4$ ($d_{xy}$
symmetry) as discussed in detail below.

In the spirit of the Hamiltonian approach 
of Ref. \cite{Cuevas1996}
the differential conductance defined as $\sigma _{ba}=dI_{b}/dV_{a}$, can be
written as \cite{Melin2003a}

\begin{equation}
\sigma _{ba}=\frac{8\pi ^{2}ep_{a}^{2}p_{b}^{2}}{h}\rho _{e,a}(\rho
_{h,b}|G_{ba,12}^{r}(eV)|^{2}-\rho _{e,b}|G_{ba,11}^{r}(eV)|^{2}),
\label{Cond_ab}
\end{equation}%
where $\rho _{e(h),a(b)}$ is the local density of states of the
electron(hole) in the lead $a(b)$, while $p_{a}$ and $p_{b}$ denote the
corresponding hopping parameters coupling the superconductor to the leads.
The quantities $G_{ba,11}^{r}\left( eV\right) $ and $G_{ba,12}^{r}\left(
eV\right) $ are the non-local propagators in the
superconducting region (indexes $1,2$ refer to electrons
and holes in Nambu space). The
first term on the right hand size of Eq. (\ref{Cond_ab}) 
is due to CAR processes, while the
second term corresponds to EC. Therefore the crossed differential
conductance is positive if CAR dominates over EC or negative in the opposite
case. In order to make contact with possible experiments  
it also convenient to analyze the
non-local resistance $R_{ba}$, given by $-\sigma_{ba}/(\sigma_{aa}%
\sigma_{bb} - \sigma_{ab}\sigma_{ba})$, where $\sigma_{aa(bb)}$ are the
local conductances which can be obtained within the same formalism \cite%
{Cuevas1996,Melin2003a}. The propagators $G^r_{\alpha\beta,ij}$ of the
coupled system are then given by
$\check{G}^{r}(E)=\left( \check{g}^{r}(E)^{-1}+i\check{\Gamma}\right) ^{-1}$,
where $\check{g}^{r}$ is the retarded Green function of the uncoupled
superconductor and $\check{\Gamma}_{\alpha \beta ,ij}=p_{\alpha }^{2}\pi
\rho _{N}\delta _{\alpha ,\beta }\delta _{i,j}$.
We have assumed that the densities of states of the normal metals are energy
independent, i.e. $\rho _{e,a(b)}=\rho _{h,a(b)}\equiv \rho _{N}$. The
symbol $^{\vee }$ here denotes $4\times 4$ matrices defined in the electrodes 
$\oplus $ Nambu space, while we reserve the symbol $^{\wedge }$ for the
reduced $2\times 2$ Nambu space. To calculate $\check{g}$ we first determine
the superconductor surface Green function 
in momentum representation, $\hat{g}_{S}(E,k_{y})$%
, using the asymptotic solutions of the Bogoliubov de Gennes equation \cite%
{Herrera}, which yields

\begin{equation}
\hat{g}_{S}^{r}(E,k_{y})=\frac{-2mi}{\hbar ^{2}D}\left( 
\begin{array}{cc}
\frac{1}{k^{-}}+\frac{\Gamma ^{2}e^{-i\Delta \varphi }}{k^{+}} & e^{i\varphi
_{-}}\left( \frac{\Gamma }{k_{1}}+\frac{\Gamma \delta }{k_{2}}\right)  \\ 
e^{\text{-}i\varphi _{+}}\left( \frac{\Gamma }{k_{1}}-\frac{\Gamma \delta }{%
k_{2}}\right)  & \frac{1}{k^{+}}+\frac{\Gamma ^{2}e^{-i\Delta \varphi }}{%
k^{-}}%
\end{array}%
\right)   \label{g_sup general}
\end{equation}%
where 
\begin{eqnarray}
k_{\pm } &=&\sqrt{k_{xF}^{2}\pm 2m\Omega /\hbar ^{2}},\;\Gamma =|\Delta
_{+}|/(E+\Omega )  \notag \\
\varphi _{\pm } &=&\arg (\Delta _{\pm }),\;\Delta \varphi =\varphi
_{+}-\varphi _{-},\;k_{xF}^{2}=k_{F}^{2}-k_{y}^{2}  \notag \\
D &=&\left( 1-\Gamma ^{2}e^{-i\Delta \varphi }\right) ,\;\Omega =\sqrt{%
E^{2}-|\Delta _{+}|^{2}} \\
\delta  &=&D(1-e^{i\Delta \varphi })/(2-2\Gamma ^{2})  \notag \\
k_{1}^{-1} &=&k_{+}^{-1}+k_{-}^{-1},\;k_{2}^{-1}=k_{+}^{-1}-k_{-}^{-1}. 
\notag
\end{eqnarray}%
In the above equations $\Delta$ is the pair potential which depends on the
wave vector, taking the values $\Delta _{+}$ and $\Delta _{-}$ along the
directions $\theta$ and $\pi -\theta$ respectively, 
where $\theta =\tan ^{-1}(k_{y}/k_{xF})$. These are given by $\Delta _{\pm
}(\theta )=\Delta _{0},$ for $s$ symmetry and $\Delta _{\pm }=\Delta
_{0}\cos (2\left( \theta \mp \alpha \right) )$ for $d$-symmetry. The retarded component is obtained by adding a small positive imaginary part $i\eta$ to the energy.
From $\hat{g}%
_{S}(E,k_{y})$ one then obtains the non-local components $\hat{g}^r_{ba}$ by

\begin{equation}
\hat{g}_{ba}^{r}(E)=\int_{-\infty }^{\infty }\hat{g}%
_{S}^{r}(E,k_{y})|f(k_{y})|^{2}e^{-ik_{y}d}dk_{y},  \label{gEyy}
\end{equation}%
where the weighting factor $f(k_{y})$, proportional to the perpendicular
wave vector $k_{xF}$, provides the appropriate connection between the
continuous model used to describe the superconducting region and the
discrete Hamiltonian approach used to obtain Eq. (\ref{Cond_ab}) 
(see Refs. \cite{Bardeen1961,Prada2004}).

As a first test of the model one can check that in the case of s-symmetry for $E<\Delta _{0}$ and 
$k_{F}d>>1$, $\sigma_{ba}\propto e^{-2d/\pi \xi }\left( \cos ^{2}(k_{F}d)-\sin
^{2}(k_{F}d)\right) /d^{3}$ with $\xi (E)=\xi _{0}/$Re$(\sqrt{1-E^{2}/\Delta ^{2}})$ and $\xi _{0}=\hbar v_{F}/(\pi \Delta _{0})$, a result which agrees with Refs. \cite{Prada2004,Maekawa2003}. 
Therefore $\sigma _{ba}$ exhibits changes in
sign on the $\lambda _{F}$ scale and its spatial average is zero \cite{Falci2001,Melin2004}.

\subsection{Results for $d_{x^2-y^2}$ symmetry}
\begin{figure}
\begin{center}
\includegraphics [scale=0.8]{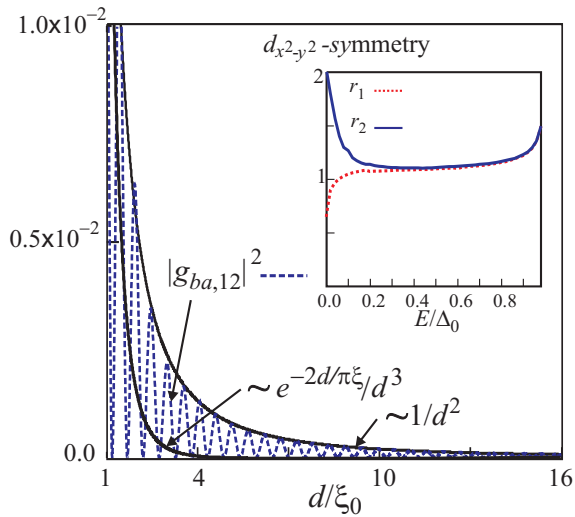}
\end{center}
\caption{(Color online) The anomalous propagator $|g_{ba,12}|^{2}$ for $d_{x^2-y^2}$
symmetry at zero energy as a function of the separation between the leads.
The propagator has been normalized to its value in the normal state at $d=0$.
The corresponding electron propagator $g_{ba,11}$ is negligible within this
scale. The envelope curve, decaying as $1/d^2$ is indicated by a full line.
The corresponding curve for $s$-symmetry with the same choice of parameters, exhibiting an exponential decay, is also
represented for comparison. The inset shows the exponents $r_{1,2}$ in the
decay laws of Eq. (\protect\ref{gdx2y2}) as a function of energy.}
\label{Fig:dx2y2}
\end{figure}

We now consider the $d_{x^{2}-y^{2}}$ symmetry. Due to the anisotropy
of the pair potential
an incoming electron from lead $a$ is scattered as
a quasiparticle in the superconductor, exploring regions where $%
E>\Delta (\theta)$ and $E<\Delta (\theta)$, with an effective
coherence length $\xi (E,\theta )=\xi _{0}/$Re$(\sqrt{1-E^{2}/\Delta (\theta
)^{2}})$ which takes values from $\xi _{0}$ to $\infty $. For this reason one typically
finds that the propagators exhibit a slower decay with distance than in the
case of $s$-symmetry. In this paper we have fixed $\Delta _{0}\sim 20$meV 
\cite{Fischer1} and $\Delta _{0}/E_{F}\sim 10^{-1}$ as typical values
for HTcS \cite{Golubov}. We also take $\eta \sim 0.002\Delta_0$ to simulate the effect of weak disorder \cite{relax}. Figure \ref{Fig:dx2y2} illustrates the
spatial dependence of the Green functions. Due to the dependence on $k_{y}$
of \ the pair potential, it is not possible to obtain an analytical
expression of their variation with $d$ as in the case
of $s$ symmetry. However, from numerical regressions for $k_{F}d>>1$, 
$|g_{ba,11}|^{2}$ and $|g_{ba,12}|^{2}$ can be fitted as

\begin{equation}
|g_{ba,11(12)}^{r}(E)|^{2} \simeq \frac{c_{1(2)}+d_{1(2)}\cos ^{2}(kd)}{%
|k_{F}d|^{r_{1(2)}}}.  \label{gdx2y2}
\end{equation}%
The values of the exponents $r_{1(2)}$ fixing the spatial decay 
are shown in the inset of Fig.\ref%
{Fig:dx2y2} as a function of energy.
For low energies ($E<<\Delta _{0}$) $k \sim k_F$, $c_{1(2)}<<d_{1(2)}$ and $%
d_{2}>>d_{1}\rightarrow 0$ for $E\rightarrow 0$, and therefore the propagator $|g_{ba,12}^{r}|$
takes a much larger value than $|g_{ba,11}^{r}|$, yielding a clear dominance
of CAR over EC in the tunnel limit.
Notice that the low energy excitations at $\theta \sim \pi /4$ give
a negligible contribution to $|g_{ba,11}^{r}|$ due to the
weighting factor in Eq. (\ref{gEyy}) \thinspace\ that is maximum at low
angles. In contrast, most of the weight in $|g_{ba,12}^{r}|$ comes from 
$\theta \sim 0$ where $\Delta$ reaches a maximum.
On the other hand, for energies higher than $E\sim 0.1 \Delta _{0}$ 
$|g_{ba,12}^{r}|$ and $|g_{ba,11}^{r}|$
tend to have the same magnitude on average.

\begin{figure}
\begin{center}
\includegraphics [scale=1.10]{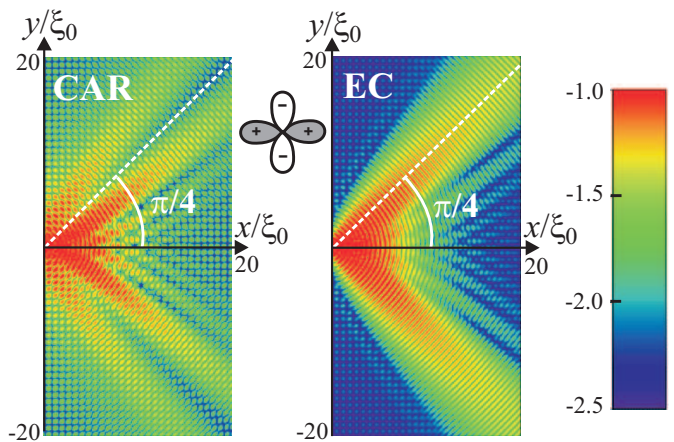}
\end{center}
\caption{(Color online) Plots of the CAR and EC contributions to the non-local conductance for the $d_{x^2-y^2}$ symmetry as one of the contacts moves inside the superconductor while the other remains fixed at $x=y=0$. The contacts are in the tunnel limit and the voltage is set to zero. Both contributions are plotted in a logarithmic scale normalized to their maximum value.}
\label{Fig:colormap}
\end{figure}

Fig. \ref{Fig:colormap} further illustrates the different behavior of CAR and EC  contributions to the non-local conductance for this orientation as one of the leads moves inside the superconductor while the second is located at the surface. These maps clearly correspond to a difraction pattern for electrons injected at one point in the surface. In spite of its complex structure one can identify the region of low angles from the surface ($\pi/2 >\theta \gtrsim \pi/4$) where CAR processes have a clear dominance and the nodal lines ($\theta \simeq \pi/4$) around which EC processes are favoured. 

\begin{figure}
\begin{center}
\includegraphics [scale=0.95]{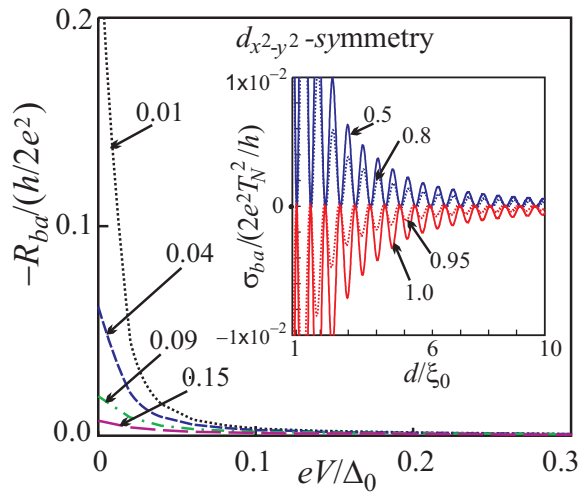}
\end{center}
\caption{(Color online) Spatial averaged non-local resistance at $d=10\protect\xi _{0}$ for $%
d_{x^{2}-y^{2}}$ symmetry as a function of the voltage for different values of the
transmission $T_{N}$. The inset shows the non-local conductance at $eV=0$ as
function of $d$ and for different values of $T_{N}$. For $T_{N}\approx 0.9$
there is a sign change in $\protect\sigma _{ba}$.}
\label{Fig:resistancedx2y2}
\end{figure}

The results for $\sigma_{ba}$ and $R_{ba}$ in the $d_{x^2-y^2}$ 
orientation for arbitrary contact transmission
are illustrated in Fig. \ref{Fig:resistancedx2y2}. It is found that an increase
in transmission leads to a reduction of the
CAR contribution while the EC one increases. 
As a consequence both quantities exhibit a change of sign when
the coupling to the leads increases.
This is illustrated for $\sigma_{ba}$ in inset of 
Fig. \ref{Fig:resistancedx2y2}. 
We can obtain further insight on this
effect at low energies where Eq. (\ref{Cond_ab}) can be approximated as
\begin{equation}
\sigma _{ba}(d)\simeq \frac{\left\vert 1-P^{4}g_{aa,12}^{r2}\right\vert
^{2}-4P^{4}\left\vert g_{aa,12}^{r}\right\vert ^{2}}{\left\vert
1+P^{4}g_{aa,12}^{r2}\right\vert ^{4}}|g_{ba,12}^{r}(d)|^{2}.
\end{equation}%
In obtaining this expression we have assumed symmetrical 
contacts ($p_{a}=p_{b}=p$) with $P=p\pi
\rho _{N}$ being the normalized hopping parameter, such that the normal
transmission for a single contact is $T_{N}=4P^{2}/(1+P^{2})^{2}$. 
Within this approximation the dependence with the separation
between the leads does not change when the transmission is increased, as it is
seen in the inset of Fig. \ref{Fig:resistancedx2y2}. 
On the other hand, this equation predicts a change in sign of $\sigma_{ba}$ 
for $P\simeq 0.72$ $(T_{N}\simeq 0.9)$ in agreement with
the numerical results in the inset of Fig \ref{Fig:resistancedx2y2}. The 
non-local resistance $R_{ba}$ averaged on a range $\sim \lambda_F$
is shown in Fig. \ref{Fig:resistancedx2y2} for $d=10\xi _{0}$. We
observe that in the low transmission regime this quantity is negative,
as it corresponds to the dominance of CAR processes, and
exhibits a peak at low bias.
\subsection{Results for $d_{xy}$ symmetry}
\begin{figure}
\begin{center}
\includegraphics [scale=0.8]{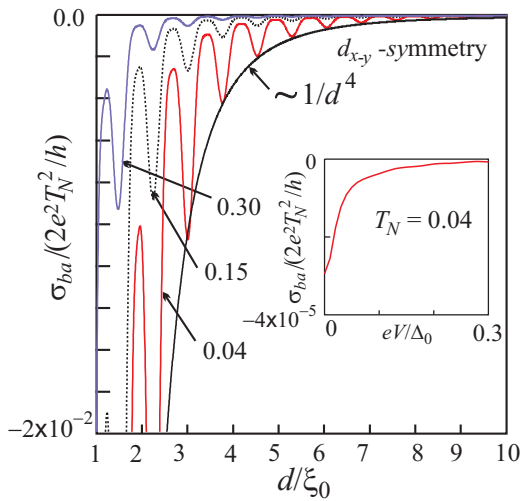}
\end{center}
\caption{(Color online) Non-local conductance at $eV=0$ for $d_{xy}$ symmetry as a function
of the distance between the leads for $T_{N}=0.04,0.15$ and $0.3$. The inset
shows the spatial averaged non-local conductance at $d=10\protect\xi _{0}$ and $%
T_{N}=0.04$ as function of the voltage.}
\label{Fig:dxy}
\end{figure}

The results \ for $d_{xy}$-symmetry $\left( \alpha =\pi /4\right) $ are
shown in Fig. \ref{Fig:dxy}. The main effect for this symmetry is the
appearance of a zero energy bound state \cite{Kashiwaya1}, which is
associated with a $1/E$ dependence in $g_{ba,11}^{r}$. On the other hand,
the distance dependence for low energy and $k_{F}d>>1$ is in this case
approximately $1/d^{4}$ both for CAR and EC processes. In this orientation, the local
Andreev reflection is zero because of diffraction of quasiparticles in the
contact \cite{Takagaki,Herrera2005} and the CAR contribution to $\sigma_{ba}$
is not zero, but is always
smaller than the EC one, leading to a negative non-local conductance as shown in Fig. \ref{Fig:dxy}. Basically, the
dominance of the EC contribution is caused by the suppression of the pair
potential along the $\theta =0$ line. The effect of varying the contact
transmission can be understood analytically within a similar approximation
as done for the $d_{x^2-y^2}$ case, which allows to obtain the following
expression for the crossed differential conductance

\begin{equation}
\sigma _{ba}(d)\simeq \frac{\left\vert g_{ba,12}^{r}\left( d\right) \right\vert
^{2}-\left\vert g_{ba,11}^{r}\left( d\right) \right\vert ^{2}}{\left\vert
1+iP^{2}g_{aa,11}^{r}\right\vert ^{4}}.
\end{equation}%
Notice that CAR and EC contributions are equally affected by the
coupling to the leads (through the $P$-dependent common denominator)
and therefore EC dominates over CAR for the whole
transmission range. 
The spatially averaged non-local conductance is 
negative and presents a zero bias peak that decreases with increasing transmission.

\section{Conclusions}
In summary we have analyzed the behavior of the crossed differential
conductance in $d$-wave superconductors in a multiterminal configuration. We
have shown that correlations between different leads exhibit an algebraical
decay instead of the exponential behavior which is typically found in
conventional superconductors. In the case of $d_{x^{2}-y^{2}}$ orientation
crossed Andreev processes are favored at low voltages and contact
transmissions, while for the $d_{xy}$ case a zero bias non-local conductance
peak appears, which is dominated by elastic-cotunneling. In both cases the
spatially averaged non-local conductance is different from zero. These
properties would allow to detect non-local transport at distances several
times larger than the characteristic coherence length in these systems.

\begin{acknowledgments} We thank H. Castro for fruitful discussions. Support by DIB of the Universidad
Nacional de Colombia, the Spanish Ministerio de Ciencia e Innovacion through contract FIS2005-06255 and
the EU program Nanoforum-EULA is acknowledged.
\end{acknowledgments}


\end{document}